\def\spose#1{\hbox to 0pt{#1\hss}}
\def\lta{\mathrel{\spose{\lower 3pt\hbox{$\mathchar"218$}}
     \raise 2.0pt\hbox{$\mathchar"13C$}}}
\def\gta{\mathrel{\spose{\lower 3pt\hbox{$\mathchar"218$}}
     \raise 2.0pt\hbox{$\mathchar"13E$}}}
\def\ge{\mathrel{\spose{\lower 3pt\hbox{$-$}}
     \raise 2.0pt\hbox{$\mathchar"13E$}}}
\def\le{\mathrel{\spose{\lower 3pt\hbox{$-$}}
     \raise 2.0pt\hbox{$\mathchar"13C$}}}

\def\simlt{\lta}

\documentclass[aps,twocolumn]{revtex4}

\usepackage{graphics}

\begin{document}

\bibliographystyle{apsrev}

\title{LISA data analysis: Source identification and subtraction}
\author{Neil J. Cornish} \affiliation{Department of Physics, Montana
State University, Bozeman, MT 59717} \author{Shane L. Larson}
\affiliation{Space Radiation Laboratory, California Institute of
Technology, Pasadena, CA 91125}

\begin{abstract}
The Laser Interferometer Space Antenna (LISA) will operate as an
AM/FM receiver for gravitational waves. For
binary systems, the source location, orientation and orbital phase are
encoded in the amplitude and frequency modulation.  The same
modulations spread a monochromatic signal over a range of frequencies,
making it difficult to identify individual sources.  We present a
method for detecting and subtracting individual binary signals from a
data stream with many overlapping signals.

\end{abstract}

\maketitle

\section{Introduction}

Estimates of the low frequency gravitational wave background below
$\sim 3$ mHz \cite{HilsBender,HBW} have suggested that the profusion
of binary stars in the galaxy will be a significant source of noise
for space based gravitational wave observatories like LISA
\cite{LPPA}.  Most of these binary sources are expected to be
monochromatic, evolving very little over the lifetime of the LISA
mission; they will thus be ever present in the data stream, and data
analysis techniques will need to be developed to deal with them.

Below $\sim 3$ mHz, it is expected that there will be more than one
binary contributing to the gravitational wave background in a given
frequency resolution bin.  Predictions suggest that the population of
binaries will be so large as to produce a confusion limited background
which will effectively limit the performance of the instrument.  In
this regime, it is likely that time delay interferometry techniques
can be employed to characterize the background \cite{HoganBender}.  At
higher frequencies, open bins appear and individual galactic binaries
(in principle) become resolvable as single monochromatic lines in the
Fourier record (a ``binary forest'').  Complications arise, however,
from the orbital motion of the LISA detector, which will modulate the
signal from an individual source, spreading the signal over many
frequency bins.  A quick method for demodulating the effect of the
orbital motion on continuous gravitational wave sources has recently
been demonstrated \cite{CLdemod}.

Unlike sources for ground based observatories, the gravitational waves
from low frequency galactic binaries are expected to be well
understood.  In principle, it should be possible to use knowledge
about the expected gravitational wave signals to ``subtract''
individual sources out of the LISA data stream, both at high
frequencies where individual sources are resolvable and at lower
frequencies where single bright sources will stand out above the rms
level of the confusion background.  The ability to perform binary
subtraction in LISA data analysis is particularly important in the
regime of the LISA floor (from $\sim 3$ mHz to the LISA
transfer frequency, $f_{*} = c/(2 \pi L) \sim 10$ mHz),
where an overlapping population of galactic binaries will severely
limit our ability to detect and study gravitational waves from other
sources, such as the extreme mass ratio inspiral of compact objects
into supermassive black holes \cite{HughesEMR,GHK}.

As will be seen, the problem of subtracting a binary out of the data
stream is intimately tied to the problem of source identification,
which is complicated by the motion of the LISA detector. Several
authors \cite{Cutler98,MooreHellings} have previously examined the
angular resolution of the observatory as a function of the time
dependent orientation. The binary subtraction problem has received
some attention in the past\cite{tuck}, but the work was not
published.

This paper examines the problem of binary subtraction using a variant of
the CLEAN algorithm \cite{Hoegbom} from electromagnetic astronomy as a
model for the subtraction procedure. The CLEAN algorithm may be concisely
described in a few steps:
\begin{list}{$\bullet$}{}
\item{Identify the brightest source in the data.}
\item{Using a model of the instrument's response function, subtract a
small portion of signal out of the data, centered on the bright source.}
\item{Remember how much was subtracted and where.}
\item{Iterate the first three steps until some prescribed level in
the data is reached.}
\item{From the stored record of subtractions, rebuild individual
sources\footnote{Electromagnetic astronomers call the product of this
step the {\it CLEAN map}.}.}
\end{list}
The implementation of the CLEAN algorithm in this paper is built
around a search through a multidimensional template space which
covers a binary source's frequency and amplitude, sky position,
inclination, polarization and orbital phase.

The format of this paper will be as follows: Section \ref{sec:Modulation}
describes the modulation  of gravitational wave signals by the motion of the
LISA detector with  respect to the sky.  Sections \ref{binarysources} and
\ref{binarymod} outline the description of the binaries used in this work,
and the effect of the detector motion on their signals.  Section \ref{templateoverlap}
describes the template space used to implement the gravitational wave CLEAN
(``gCLEAN'') algorithm.  Section \ref{noise} reviews the expected contributions
of instrumental noise and the effects on the data analysis procedure.
Section \ref{gclean} describes and demonstrates the gCLEAN procedure in detail.
Lastly, a discussion of outstanding problems and future work is given
in Section \ref{future}.

\section{Signal Modulation}\label{sec:Modulation}

LISA's orbital motion around the Sun introduces amplitude, frequency
and phase modulation into the observed gravitational wave signal.  The
amplitude modulation results from the detector's antenna pattern being
swept across the sky, the frequency modulation is due to the Doppler
shift from the relative motion of the detector and source, and the
phase modulation results from the detector's varying response to the
two gravitational wave polarizations.  The general expression
describing the strain measured by the LISA detector is quite
complicated\cite{cr}, but we need only consider low frequency,
monochromatic plane waves.  Here low frequency is defined relative to
the transfer frequency\cite{cl} of the LISA detector, $f_*\approx
10$ mHz.  The low frequency LISA response function was first derived
by Cutler\cite{Cutler98}, but we shall use the simpler description
given in Ref.~\cite{cr}.

A monochromatic plane wave propagating in the $\widehat \Omega$
direction can be decomposed:
\begin{equation}
   {\bf h}(t,f) = A_+ \cos(2\pi f t+\varphi_0)
   {\mbox{\boldmath$\epsilon$}}^{+} + A_\times \sin(2\pi f
   t+\varphi_0) {\mbox{\boldmath$\epsilon$}}^{\times},
   \label{gwh}
\end{equation}
where $A_+$ and $A_\times$ are the amplitudes of the two polarization
states and
\begin{eqnarray}
   {\mbox{\boldmath$\epsilon$}}^{+} &=& \hat{p}\otimes\hat{p} -
   \hat{q}\otimes\hat{q} , \nonumber \\
   {\mbox{\boldmath$\epsilon$}}^{\times} &=& \hat{p}\otimes\hat{q} +
   \hat{q}\otimes\hat{p} \, ,
   \label{polarizeEpsilon}
\end{eqnarray}
are polarization tensors.  Here $\hat{p}$ and $\hat{q}$ are vectors
that point along the principal axes of the gravitational wave.  For a
source located in the $\hat{n}=-\widehat \Omega$ direction described
by the ecliptic coordinates $(\theta,\phi)$ we can construct the
orthogonal triad
\begin{eqnarray}\label{wave}
   &&\hat{u} = \cos\theta\cos\phi \, \hat{x}
   +\cos\theta\sin\phi \, \hat{y} -\sin\theta \, \hat{z}
   \nonumber \\
   &&\hat{v} = \sin\phi \, \hat{x} -\cos\phi \, \hat{y} \nonumber
   \\
   &&\hat{n} = \sin\theta \cos\phi \, \hat{x} + \sin
   \theta\sin\phi \, \hat{y} +\cos\theta \, \hat{z}\, .
\end{eqnarray}
This allows us to write
\begin{eqnarray}\label{polten}
   {\mbox{\boldmath$\epsilon$}}^{+} &=&\cos 2\psi\, {\bf e}^{+} -
   \sin 2\psi\, {\bf e}^{\times}, \nonumber \\
   {\mbox{\boldmath$\epsilon$}}^{\times} &=& \sin 2\psi\, {\bf
   e}^{+} + \cos 2\psi\, {\bf e}^{\times},
\end{eqnarray}
where 
\begin{eqnarray}\label{eten}
   {\bf e}^{+} &=& \hat{u}\otimes \hat{u} - \hat{v}\otimes \hat{v},
   \nonumber \\
   {\bf e}^{\times} &=& \hat{u}\otimes \hat{v} + \hat{v}\otimes
   \hat{u},
\end{eqnarray}
and the polarization angle $\psi$ is defined by
\begin{equation}
   \tan \psi = -\frac{\hat{v}\cdot\hat{p}}{\hat{u}\cdot\hat{p}}.
\end{equation}
The strain produced in the detector is given by
\begin{equation}\label{sbasic}
   s(t) = A_+ F^+(t) \cos \Phi(t) + A_\times F^\times(t) \sin \Phi(t) \, ,
\end{equation}
where
\begin{equation}
    \Phi(t)=2\pi f t +\varphi_0 + \phi_D(t) \, .
\end{equation}
Here $\phi_D(t)$ describes the Doppler modulation
and $F^+(t)$, $F^\times(t)$ are the detector beam patterns
\begin{eqnarray}\label{beam}
   F^+(t) &=& \frac{1}{2}\left[\cos 2\psi \, D^+(t) - \sin 2\psi
   \, D^\times(t) \right]\nonumber\\
   F^\times(t) &=& \frac{1}{2}\left[\sin 2\psi \, D^+(t) + \cos
   2\psi \, D^\times(t) \right],
\end{eqnarray}
where
\begin{eqnarray}
   && D^+(t) = \frac{\sqrt{3}}{64} \Big[ -36 \sin^2\theta
   \sin(2\alpha(t)-2\lambda ) \nonumber\\
   && \quad + (3+\cos 2\theta)\Big(\cos 2\phi (9 \sin
   2\lambda-\sin(4\alpha(t)-2\lambda)) \nonumber\\
   && \hspace*{0.8in} +\sin 2\phi(\cos (4\alpha(t)-2\lambda) - 9\cos
   2\lambda )\Big) \nonumber\\
   && \quad - 4 \sqrt{3} \sin 2\theta
   \Big(\sin(3\alpha(t)-2\lambda-\phi)\nonumber\\
   && \hspace*{0.8in} - 3\sin(\alpha(t)-2\lambda+\phi) \Big) \Big],
\end{eqnarray}
and
\begin{eqnarray}
   D^\times(t) &=& \frac{1}{16} \Big[ \sqrt{3} \cos \theta
   \Big(9\cos (2\lambda-2\phi) \nonumber\\
   && \quad - \cos(4\alpha(t) - 2\lambda-2\phi)\Big) \nonumber\\
   &-& 6\sin\theta \Big(\cos(3\alpha(t)-2\lambda-\phi) \nonumber\\
   && \quad + 3\cos(\alpha(t)-2\lambda+\phi)\Big) \Big].
\end{eqnarray}
The quantity $\alpha(t) = 2\pi f_m t + \kappa$ describes the orbital
phase of the LISA constellation, which orbits the Sun with frequency
$f_m={\rm year}^{-1}$.  The constants $\kappa$ and $\lambda$ specify
the initial orbital phase and orientation of the detector\cite{cr}.
We set $\kappa=0$ and $\lambda=3\pi/4$ in order to reproduce the
initial conditions chosen by Cutler\cite{Cutler98}. The Doppler modulation
depends on the source location and frequency, and on the velocity of
the guiding center of the detector:
\begin{equation}
    \phi_D(t) =  2\pi f \frac{R}{c}\sin\theta \cos(2\pi f_m t - \phi)
\end{equation}
Here $R$ is the separation of the detector from the barycenter, so
$R/c$ is the light travel time from the guiding center of the detector
to the barycenter.

The expression for the strain in the detector can be re-arranged using
double angle identities to read:
\begin{equation}
    s(t) = A(t) \cos \Psi(t)
\end{equation}
where
\begin{equation}
    \Psi(t)=2\pi f t +\varphi_0 + \phi_D(t) + \phi_P(t)\, .
\end{equation}
The amplitude modulation $A(t)$ and
phase modulation $\phi_P(t)$ are given by
\begin{eqnarray}
    A(t) &=& \left[ (A_+ F^+(t))^2+(A_\times F^\times (t))^2
    \right]^{1/2} \\
    \nonumber \\
    \phi_P(t) & = & - {\rm arctan}\left(\frac{A_\times
    F^\times(t)}{A_+ F^+(t)}\right) \, .
\end{eqnarray}
Each of the modulation functions are periodic in
harmonics of $f_m$. To get a feel for how each
modulation affects the signal, we begin by turning off all but one
modulation at a time and look at how each individual term affects the
signal.

\subsection{Amplitude modulation}
Amplitude modulation derives from the sweep of the detector's antenna
pattern across the sky due to the observatory's
orbital motion, which for LISA gives a modulation frequency, $f_{m} =
1/{\rm year}$.  Pure amplitude modulation takes the form
\begin{equation}
   s(t) = A(t) \cos(2 \pi f t+\varphi_0) \, .
   \label{amplitudeSignal}
\end{equation}
The amplitude, $A(t)$ is modulated by the orbital motion, and may
be expanded in a Fourier series:
\begin{equation}
   A(t) = \sum_{n=-\infty}^{\infty} a_n e^{2\pi i f_m n t}
\end{equation}
which allows the signal in Eq.\ (\ref{amplitudeSignal}) to be written
\begin{equation}
   s(t) = \Re \left( \sum_{n=-\infty}^{\infty} a_n e^{2\pi i (f+f_m n)
   t} e^{i\varphi_0}\right) \, .
\end{equation}
Thus, the Fourier power spectrum of $s(t)$ will have sidebands about
the carrier frequency $f$ of the signal, spaced by the modulating
frequency $f_m$.  The bandwidth, $B$, of the signal is defined to be
the frequency interval which contains 98\% of the total power:
\begin{equation}
   B = 2N f_m \, ,
   \label{ampBand}
\end{equation}
where $N$ is determined empirically by
\begin{equation}
   \sum_{n=-N}^{N} \vert a_n \vert^2 \geq 0.98
   \sum_{n=-\infty}^{\infty} \vert a_n \vert^2 \, .
\end{equation}
Typical LISA sources give rise to an amplitude modulation with $N=2$
and using Eq.\ (\ref{ampBand}) a bandwidth of $B = 4f_{m} = 1.3 \times 10^{-4}$ mHz.

\subsection{Frequency modulation}
Doppler (frequency) modulation of signals occurs because of relative
motion between the detector and the source, and depends on the angle
between the wave propagation direction $\widehat{\Omega}$ and
the velocity vector of the guiding center. Pure Doppler modulation takes
the form
\begin{equation}
   s(t) = A \cos\left[2 \pi f t + \beta \cos(2\pi f_m t+\delta)
   +\varphi_0 \right] \, ,
   \label{DopplerModSignal}
\end{equation}
where $\beta$ and $\delta$ are constants.
Using the Jacobi-Anger identity to write the Fourier expansion as
\begin{equation}
   e^{i\beta \sin(2\pi f_m t)} = \sum_{n=-\infty}^{\infty} J_n(\beta)
   e^{2\pi i f_m n t}
\end{equation}
allows the signal in Eq.\ (\ref{DopplerModSignal}) to be written
\begin{equation}
   s(t) = \Re\left( A \sum_{n=-\infty}^{\infty} J_n(\beta) e^{2\pi i
   (f+f_m n) t} e^{i \varphi_0}e^{i n (\delta+\pi/2)}\right)\, .
\end{equation}
Once again, the Fourier power spectrum of $s(t)$ will have sidebands
about the carrier frequency $f$, spaced by the modulating frequency
$f_m$.  The bandwidth of the signal is given by
\begin{equation}
   B = 2(1+\beta) f_m
\end{equation}
For LISA, the parameter $\beta$ (called the {\it modulation index}),
which encodes the description of the detector motion relative to the
source, is given by
\begin{equation}
   \beta = 2\pi f \frac{R}{c} \sin\theta \, .
\end{equation}
Sources in the equatorial plane have bandwidths ranging from $B = 2.6
\times 10^{-4}$ mHz at $f=1$ mHz to $B = 2.1 \times 10^{-3}$ mHz at
$f=10$ mHz.

\subsection{Phase modulation}
Phase modulation is a consequence of the fact that the detector has
different sensitivities to the two gravitational wave polarization
states, $+$ and $\times$, characterized by the two detector beam
patterns, $F^{+}(t)$ and $F^{\times}(t)$.  The variation of these beam
patterns is a function of the detector motion (see Eq.\
(\ref{beam})), and modulates the phase.  Phase modulation
takes a similar form to the frequency modulation.  Expanding
$\phi_P(t)$ in a Fourier sine series yields a signal
\begin{equation}
    s(t) = A \cos(2 \pi f t + \varphi_0+\sum_n \beta_n \sin(2\pi f_m n
    t+ \delta_n) ) \, .
\end{equation}
Again, the the Fourier power spectrum of $s(t)$ has sidebands about
the carrier frequency $f$, spaced by the modulating frequency $f_m$.
The main difference is that the Fourier amplitude of the $k^{\rm th}$
sideband (located at $f+ k f_m$) is given by
\begin{equation}
    c_k = A \prod_n \sum_{l_n} J_{l_n}(\beta_n) e^{i l_n \delta_n}
    e^{i \varphi_0} \quad {\rm where} \quad k=\sum_n l_n \, .
\end{equation}
Since the $\beta_n$'s for LISA are independent of frequency (at least
in the low frequency approximation used here), the bandpass of
the phase modulated signal is independent of the carrier frequency.
Empirically we find the bandwidth $B \approx 10^{-4}$ mHz.

\subsection{Total modulation}\label{subsec:totalModulation}
It is possible to combine the amplitude, frequency and phase
modulations together to arrive at an analytic expression for the full
signal modulation.  The carrier frequency $f$ develops sidebands spaced
by the modulation frequency $f_m$.  The total modulation is most
easily computed beginning from (\ref{sbasic}). One can write
\begin{eqnarray}
   F^+ &=& \sum_{n=-4}^4 p_n e^{2\pi i f_m n t} \nonumber \\
   \nonumber \\
   F^\times &=& \sum_{n=-4}^4 c_n e^{2\pi i f_m n t} \, ,
\end{eqnarray}
where the small number of non-zero Fourier coefficients can be
attributed to the quadrupole approximation for the beam pattern.
The coefficients $p_n$ and $c_n$ can be read off from (\ref{beam})
in terms of $( \theta,\,  \phi)$ and $\psi$.

Our next task is to Fourier expand $\cos\Phi(t)$ and $\sin\Phi(t)$,
being careful to take into account the fact that we are doing finite
time Fourier transforms, so $f$ will not be an integer multiple of
$f_m$.  In other words, writing
\begin{equation}
    e^{2\pi i f t} = \sum_{n=-N}^{N} a_n e^{2\pi i f_m n t} \, ,
\end{equation}
we find in the limit $N \gg 1$ that
\begin{equation}
    a_n \simeq {\rm sinc}(\pi x_n) e^{i \pi x_n} \quad {\rm where}
    \quad x_n= \frac{f}{f_m} - n \, .
\end{equation}
The coefficients are highly peaked about $n={\rm int}(f/f_m)$, where
the function ``int'' returns the nearest integer to its argument.
The maximum bandwidth occurs when the
remainder $f/f_m-n$ equals $1/2$; the max bandwidth is equal to $20
f_m$ for 98\% power ($36 f_m$ for 99\% power).  Putting everything
together we find
\begin{eqnarray}
    F^+ \cos\Phi(t) &=& \Re \biggl( \left[ \sum_{k} J_k(\beta)
    e^{2\pi i f_m k t} e^{i k (\pi/2-\phi)} \right] \nonumber \\
    && \hspace*{-0.7in} \times e^{i\varphi_0} \left[ \sum_{l} p_l
    e^{2\pi i f_m l t} \right] \left[ \sum_{n} a_n e^{2\pi i f_m n t}
    \right] \biggr).
\end{eqnarray}
and
\begin{eqnarray}
    F^\times \sin\Phi(t) &=& \Im \biggl( \left[ \sum_{k}
    J_k(\beta) e^{2\pi i f_m k t} e^{i k (\pi/2-\phi)} \right]
    \nonumber \\
    && \hspace*{-0.7in} \times e^{i\varphi_0} \left[ \sum_{l} c_l
    e^{2\pi i f_m l t} \right] \left[ \sum_{n} a_n e^{2\pi i f_m n t}
    \right] \biggr).
\end{eqnarray}
It follows that the Fourier expansion of $s(t)$ is described by the
triple sum
\begin{equation}\label{quick}
    s_q = \frac{1}{2} e^{i\varphi_0} \sum_l (A_+ p_l + e^{i 3\pi/2}A_\times c_l)
    \sum_n a_n \sum_k J_k(\beta)\, ,
\end{equation}
where $q=k+l+n$.  The limited bandwidth of the various modulations
allows us to restrict the sums: $-(1+\beta) \leq k \leq (1+\beta)$,
$-4 \leq l \leq 4$ and $-10 < n - {\rm int}(f/f_m)< 10$.  Using
(\ref{quick}) we can compute the discrete Fourier transform of $s(t)$
very efficiently.

The source identification and subtraction scheme used in this work
depends on the development and use of a template bank covering
a large parameter space.  As such, issues related to efficient
computing are of interest in order to make the problem tractable in a
reasonable amount of time.  A number of simplifying factors allow the
problem to be compactified significantly, with great savings in
computational efficiency.

The quantities $p_n$ and $c_n$ only depend on $\theta,\, \phi,$
and $\psi$, so they can be pre-computed and stored.  The complete
template bank can then be built using (\ref{quick}) by stepping
through a grid in $f$, $\varphi_0$ and the ratio $A_\times/A_+$.
The computational saving as
compared to directly generating $s(t)$ for each of the six search
parameters is a factor of $\sim 10^5$ in computer time.

Another big saving in computer time is based on the following
observation: The Fourier expansions of sources $a$ and $b$ with all
parameters equal save their frequencies, which differ by an integer
multiple, $m$, of the modulation frequency $f_m$, are related:
\begin{equation}
     s^a_{q}- s^b_{q+m} \simeq \pi m f_m \frac{R}{c}\sin\theta (
     s^a_{q+1}-s^a_{q-1} ) \, .
\end{equation}
Thus, so long as $m \simlt 10^4$, we have $s^a_{q} \approx s^b_{q+m}$.
This allows us to use a set of templates generated at a frequency $f$
to cover frequencies between $f\pm 10^4 f_m$. These savings
mean that our Fourier space approach to calculating the template
bank are a factor of $10^9$ times faster than a direct computation
in the time domain!

\section{Binary Sources}\label{binarysources}

With the exception of systems that involve supermassive black holes,
all of the binary systems that can be detected by LISA are well
described by the post-Newtonian approximation to general relativity.
Most of these sources can be adequately described as circular Newtonian
binaries, and the gravitational waves they produce can be calculated
using the quadrupole approximation. In terms of these approximations,
a circular Newtonian binary produces waves propagating in the
$\widehat{\Omega}$ direction with amplitudes
\begin{eqnarray}
    A_+ &=& {\cal A} \left(1 + (\widehat{L} \cdot
    \widehat{\Omega})^2\right) \nonumber \\
    A_\times &=& 2 {\cal A} \, \widehat{L} \cdot
    \widehat{\Omega}
\end{eqnarray}
where
\begin{equation}
    {\cal A} = \frac{2 M_1 M_2}{r d}.
\end{equation}
Here $r$ is the distance between mass $M_1$ and $M_2$, $d$ is the
distance between the source and the observer, and $\widehat{L}$ is a
unit vector parallel to the binary's angular momentum vector.
The gravitational waves have frequency
\begin{equation}
    f = 2f_{\rm orb} = \frac{1}{\pi} \sqrt{\frac{M_1+M_2}{r^3}} \, .
\end{equation}
The generalization to elliptical Newtonian binaries is given in Peters and
Mathews\cite{pm}.  They found that elliptical binaries produce
gravitational waves at harmonics of the orbital frequency $f_{\rm orb}$.  For
small eccentricities, most of the power is radiated into the second
harmonic, with the portion of the power radiated into higher harmonics
increasing with increasing eccentricity.  From a data analysis
perspective, an eccentric binary looks like a collection of circular
binaries located at the same position on the sky, with frequencies
separated by multiples of $f_{\rm orb}$.  One strategy to search for eccentric
binaries would be to conduct a search for individual circular
binaries, then check to see if binaries at a certain location form
part of a harmonic series.  If they do, the relative amplitude of the
harmonics can be used to determine the eccentricity.

The polarization angle of a circular binary is related to its
angular momentum vector orientation, $\widehat{L} \rightarrow (\theta_L,\phi_L)$,
by \cite{cr}
\begin{equation}
   \tan \psi = \frac{\cos\theta\cos(\phi-\phi_L)\sin\theta_L
   -\cos\theta_L\sin\theta} {\sin\theta_L \sin(\phi-\phi_L)} \, ,
\end{equation}
The inclination of a circular binary, $\imath$ is given by
\begin{eqnarray}
\cos\imath &=& -\widehat{L} \cdot \widehat{\Omega} \nonumber \\
& = & \cos\theta_L \cos\theta + \sin\theta_L \sin\theta \cos(\phi_L - \phi)
\end{eqnarray}
It follows that the amplitude and phase modulation depend on four parameters. Two
are the sky position $(\theta,\phi)$ and the other two are either the
the angular momentum direction $(\theta_L,\phi_L)$, or the polarization angle $\psi$
and the inclination $\imath$. We found $(\imath,\psi)$ to be easier to work with
as the quadrupole degeneracy between sources with parameters
$(\psi,\varphi_0)$ and $(\psi+\pi/2,\varphi_0+\pi)$ is explicit in these coordinates.
The total gravitational wave
signal from a Newtonian binary depends on seven parameters: $\vec\lambda
\rightarrow (\theta,\, \phi,\, \imath,\,
\psi,\, \varphi_0,\, f,\, {\cal A})$. The parameter
space has topology $S^2 \times T^3 \times R^2$. The parameters $\theta$ and $\phi$
range over their usual intervals: $\theta \in [0,\pi]$ and $\phi \in [0,2\pi]$.
The inclination and polarization have ranges: $\imath \in [0,\pi]$ and $\psi \in [0,\pi]$.
Because of the quadrupole degeneracy discussed above, we restrict the range of the
orbital phase: $\varphi_0 \in [0,\pi)$.

\section{Binary Signal Modulation}\label{binarymod}

\begin{figure}[ht]
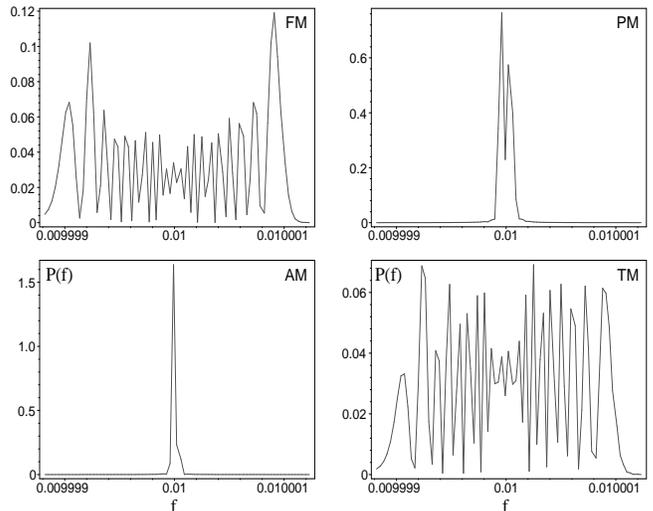

\vspace{65mm}
\includegraphics{pow_01_F.ps}
\includegraphics{pow_01_P.ps}
\includegraphics{pow_01_A.ps}
\includegraphics{pow_01_T.ps}
\caption{Power spectra showing the effects of frequency (FM), phase
(PM) and amplitude (AM) modulation separately and all together (TM).
The gravitational wave has frequency $10$ mHz.}
\label{Modulation01}
\end{figure}

\begin{figure}[ht]
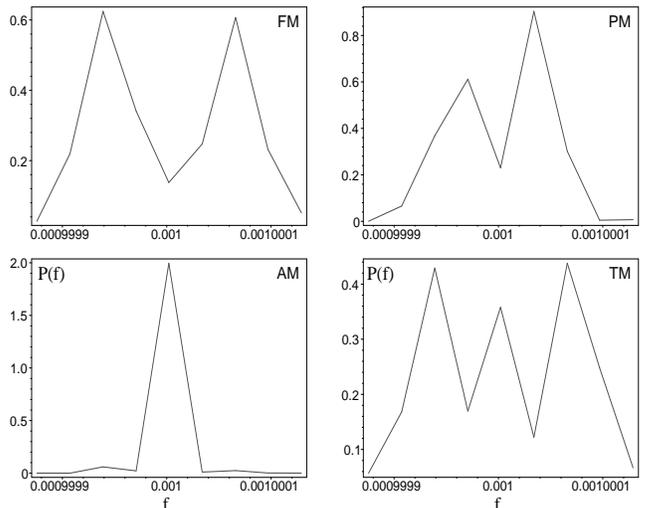

\vspace{65mm}
\includegraphics{pow_001_F.ps}
\includegraphics{pow_001_P.ps}
\includegraphics{pow_001_A.ps}
\includegraphics{pow_001_T.ps}
\caption{Power spectra showing the effects of frequency (FM), phase
(PM) and amplitude (AM) modulation separately and all together (TM).
The gravitational wave has frequency $1$ mHz.}
\label{Modulation001}
\end{figure}

The effects of amplitude, frequency and phase modulation on two binary
sources with barycenter frequencies of 10 and 1 mHz are shown in
Figures \ref{Modulation01} and \ref{Modulation001} respectively.  The
sources have all parameters equal save their frequencies, and are
located close to the galactic center.  We see that frequency
modulation dominates at 10 mHz, while frequency and phase modulation
become comparable at 1 mHz.

One of the main effects of the modulations is to spread the power
across a bandwidth $B\simeq 2(1+2\pi f \frac{R}{c} \sin\theta)
f_m$.  This, combined with LISA's antenna pattern, means that the
strain in the detector is often considerably less
than the strain of the wave. The effect can be quantified in terms of
the amplitude of the detector response, $A$, and the intrinsic amplitude
of the source $\cal{A}$. Suppose that a source is responsible for a
strain in the detector $s(t)$. Defining $A$ as the orbit-averaged response:
\begin{equation}
A^2 = \frac{1}{T} \int_0^T s^2(t) \, dt,
\end{equation}
we find from (\ref{sbasic}) that
\begin{equation}\label{amps}
A^2 = \frac{1}{2} {\cal A}^2 \left((1+\cos^2\imath)^2 \langle F_+^2 \rangle
+ 4\cos^2\imath \langle F_\times^2\rangle\right),
\end{equation}
where the orbit-averaged detector responses are given by
\begin{eqnarray}
&&\langle F_+^2\rangle = \frac{1}{4}\left( \cos^2 2\psi \langle D_+^2 \rangle
-\sin 4 \psi \langle D_+ D_\times \rangle + \sin^2 2 \psi \langle D_\times^2 \rangle
\right) \nonumber \\
&&\langle F_\times^2\rangle = \frac{1}{4}\left( \cos^2 2\psi \langle D_\times^2 \rangle
+\sin 4 \psi \langle D_+ D_\times \rangle + \sin^2 2 \psi \langle D_+^2 \rangle
\right) \nonumber \\
&&\langle F_+ F_\times\rangle = \frac{1}{8}\left( \sin 4 \psi (\langle D_+^2 \rangle
- \langle D_\times^2 \rangle) + 2\cos 4 \psi \langle D_+ D_\times \rangle \right)
\nonumber \\
\end{eqnarray}
and
\begin{eqnarray}
\langle D_+ D_\times \rangle &=& \frac{243}{512}\cos\theta \sin 2\phi (2\cos^2\phi-1)
(1+\cos^2\theta) \nonumber \\
\langle D_\times^2 \rangle &=& \frac{3}{512}\left( 120\sin^2\theta +\cos^2\theta
  +162\sin^2 2\phi \cos^2\theta \right) \nonumber \\
\langle D_+^2 \rangle &=& \frac{3}{2048}\left( 487 +158\cos^2\theta +7\cos^4\theta
\right. \nonumber \\
&& \left.  \quad -162\sin^2 2\phi(1+\cos^2\theta)^2\right)\, .
\end{eqnarray}
The relative amplitude $A/{\cal A}$ depends on the source declination $\theta$,
right ascension $\phi$, inclination $\imath$ and polarization angle $\psi$.

\begin{table}
    \centering
    \caption{Properties of the six nearest interacting white dwarf
    binaries.  Physical data from Hellier \cite{Hellier}, periods
    taken from NSSDC catalog 5509\cite{NSSDC}.  Spectral amplitudes
    are computed using the methods of this paper for one year of
    observations.  The masses quoted in units of the solar mass
    $M_{\odot}$, the orbital periods are in seconds, the distances are
    in parsecs and the strain spectral densities are in units of
    $10^{-19}$ Hz$^{-1/2}$.} \vspace*{0.1in}
    \begin{tabular}{|l|c|c|c|c|c|c|c|}
     \colrule
    Name & $\; m_{1} \;$ & $\; m_{2} \; $ &
    $\; P_{orb}\; $ &
     $\; d \; $ & $h_{f}^{B}$ & $ h_{f}^{D}$ & $h_{f}^{D} / h_{f}^{B}$  \\
      \colrule
    & & & & & & & \\
    AM CVn    & $0.5$ & $0.033$ & $1028.76$ & $100$ & $21.2$ & $2.34$ & 0.111\\
    CP Eri    & $0.6$ & $0.02$  & $1723.68$ & $200$ & $5.19$ & $1.06$ & 0.205\\
    CR Boo    & $0.6$ & $0.02$  & $1471.31$ & $100$ & $11.5$ & $1.63$ & 0.141\\
    GP Com    & $0.5$ & $0.02$  & $2791.58$ & $200$ & $3.32$ & $0.44$ & 0.133\\
    HP Lib    & $0.6$ & $0.03$  & $1118.88$ & $100$ & $20.7$ & $4.53$ & 0.219\\
    V803 Cen  & $0.6$ & $0.02$  & $1611.36$ & $100$ & $10.9$ & $1.89$ & 0.174\\
     \colrule
    \end{tabular}
    \label{BinaryTable}
\end{table}

Table 1 illustrates the power spreading effect for the six nearest
interacting white dwarf binaries. Random numbers were used for the unknown
parameters $\imath$ and $\psi$. The average strain spectral density
in the detector, $ h_{f}^{D}$, is between 5 and ten times below the
strain spectral density at the barycenter $h_{f}^{B}$.  The effect is more
significant at higher frequencies since the bandwidth increases with
frequency.

\begin{figure}[ht]
\vspace{55mm}
\includegraphics{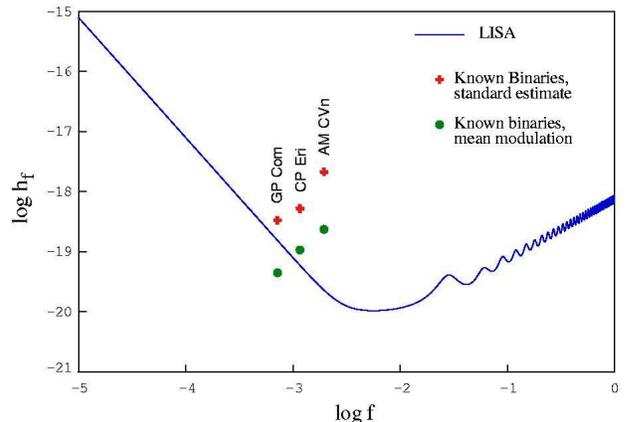}
\caption{The strain spectral densities of three nearby interacting
white dwarf binaries plotted against the standard LISA noise curve.}
\label{wdb3stars}
\end{figure}

\begin{figure}[ht]
\vspace{55mm}
\includegraphics{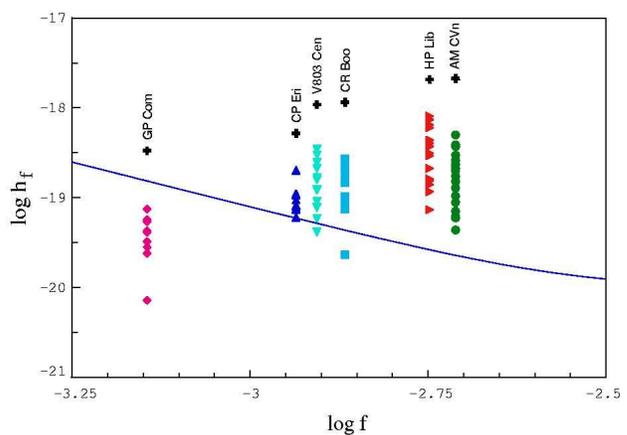}
\caption{The modulated strain spectral densities of the six nearest
interacting white dwarf binaries plotted against the standard LISA
noise curve.  Crosses mark standard estimates for the known binaries,
while alternate symbols mark modulated Fourier components.}
\label{wdbfullSpectrum}
\end{figure}

The signals from three of these binaries, averaged over their
bandwidths, are plotted against the standard LISA noise curve in
Figure \ref{wdb3stars}.  The complete signals for all six binaries are
shown in Figure \ref{wdbfullSpectrum}.  The strain spectra appear as
nearly vertical lines of dots due to the highly compressed frequency
scale.

\section{Template Overlap}\label{templateoverlap}
\subsection{Template metric}
The templates are constructed by choosing the six parameters
$\vec{\lambda} \rightarrow (f, \theta, \phi, \imath, \psi, 
\varphi_0)$ and forming the noise-free detector response corresponding
to a source with those parameters:
\begin{equation}
   s(t,\vec{\lambda}) = A(t,\vec{\lambda}) \cos \Psi(t,\vec{\lambda})
\end{equation}
We need to determine how closely the templates need to be spaced to
give a desired level of overlap. The overlap of templates with
parameters $\vec{\lambda}_1$ and $\vec{\lambda}_2$ is defined:
\begin{equation}
   R(\vec{\lambda}_1, \vec{\lambda}_2) = \frac{\langle  s(t,\vec{\lambda}_1)
   \vert s(t,\vec{\lambda}_2) \rangle}
   {\langle  s(t,\vec{\lambda}_1) \vert
   s(t,\vec{\lambda}_1) \rangle^{1/2}
    \langle  s(t,\vec{\lambda}_2) \vert
   s(t,\vec{\lambda}_2) \rangle^{1/2}}\, ,
\end{equation}
with the inner product
\begin{equation}
\langle a(t) \vert b(t) \rangle = \int_0^T a(t) b(t) \, dt .
\end{equation}
Suppose we have two templates, one
with parameters $\vec{\lambda}$ and the other with parameters
$\vec{\lambda}+\delta\vec{\lambda}$.  To leading order in
$\delta\vec{\lambda}$ the overlap is given by \cite{ben}
\begin{equation}
   R(\vec{\lambda},\vec{\lambda}+\delta\vec{\lambda})
= 1 - g_{ij}(\vec{\lambda})
                     \Delta \lambda^i\Delta\lambda^j
\end{equation}
where $g_{ij}$ is the template space metric
\begin{eqnarray}
g_{ij}(\vec{\lambda}) &=& \frac{\langle \partial_i s(t,\vec{\lambda})
\vert \partial_j s(t,\vec{\lambda})\rangle}
{2 \langle  s(t,\vec{\lambda}) \vert
   s(t,\vec{\lambda}) \rangle} \nonumber \\
&&
- \frac{\langle s(t,\vec{\lambda}) \vert
\partial_i s(t,\vec{\lambda})\rangle \langle s(t,\vec{\lambda}) \vert
\partial_j s(t,\vec{\lambda})\rangle}{2 \langle  s(t,\vec{\lambda}) \vert
   s(t,\vec{\lambda}) \rangle^2 } \, .
\end{eqnarray}
Using the fact that $\Psi(t,\vec{\lambda})$ varies much faster
than $A(t,\vec{\lambda})$ we find
\begin{eqnarray}
   g_{ij}(\vec{\lambda}) &=& \frac{\langle \partial_i A \vert
\partial_j A \rangle + \langle A \partial_i \Psi \vert
A \partial_j \Psi \rangle}{2 \langle A \vert A \rangle} \nonumber \\
&& - \frac{\langle A \vert \partial_i A \rangle \langle A \vert
\partial_j A \rangle}{2 \langle A \vert A \rangle^2} \, .
\end{eqnarray}

Ignoring the sub-dominant amplitude and phase modulation allows us to
analytically compute the ``Doppler metric''
\begin{eqnarray}\label{met}
   ds^2 &=& g_{ij}(\vec{\lambda})\Delta \lambda^i\Delta \lambda^j
  \nonumber \\
    &=& \frac{2 \pi^2}{3} T^2 df^2 + \pi T df d
   \varphi_0 + \frac{1}{2} d\varphi_0^2 \nonumber \\
   && - 2\pi f \frac{R}{c} T df \left( \cos \theta \sin \phi
   d \theta + \sin \theta \cos \phi d \phi \right) \nonumber
   \\
   && +\pi^2 f^2 \left(\frac{R}{c}\right)^2 \left( \cos^2
   \theta d \theta^2 + \sin^2 \theta d \phi^2\right).
\end{eqnarray}
Here $T= 1$ year is the observation time.
We have to go beyond the Doppler approximation to find metric
components that involve $\imath$ and $\psi$. The computations are
considerably more involved, as are the resulting expression. For
example, including all modulations we find
\begin{equation}
g_{\imath\imath} = \frac{2\langle F_+^2 \rangle\langle F_\times^2\rangle \sin^2\imath
\left(\sin^2\imath+2\cos^4\imath\right)}{\left((1+\cos^2\imath)^2 \langle F_+^2 \rangle
+ 4\cos^2\imath \langle F_\times^2\rangle\right)^2}
\end{equation}
and
\begin{eqnarray}
g_{\psi\psi} &=& \frac{2 (1+\cos^2\imath)^2 \langle F_\times^2 \rangle
+ 4\cos^2\imath \langle F_+^2\rangle}{(1+\cos^2\imath)^2 \langle F_+^2 \rangle
+ 4\cos^2\imath \langle F_\times^2\rangle} \nonumber \\
&& - \frac{2 \sin^4\imath \, \langle F_+ F_\times\rangle}{\left((1+\cos^2\imath)^2
\langle F_+^2 \rangle
+ 4\cos^2\imath \langle F_\times^2\rangle\right)^2}
\end{eqnarray}
We have been able to derived exact expressions for all the metric components, but
they are cumbersome and not very informative. For most
purposes the simple Doppler metric is sufficient.

\subsection{Overlap of parameters}\label{subsec:OverlapParams}
An important application of the Doppler metric in Eq.\ (\ref{met}) is
the determination of parameter overlap, which has great bearing on the
placement of templates. In regions where large
variations of the overlap can be seen for small changes in parameters,
templates must be spaced closely to distinguish between different
realizable physical situations.  In regions where the change in
overlap is small for small changes in parameters, the templates can be
spaced more widely.  The Doppler metric depends on only four parameters.
Taking constant slices through the parameter space for any two of the
four will produce a metric which can be used to plot level curves of
the overlap function $R(\vec{\lambda}_1, \vec{\lambda}_2)$
as a function of two parameters.

\begin{figure}[ht]
\vspace{60mm}
\includegraphics{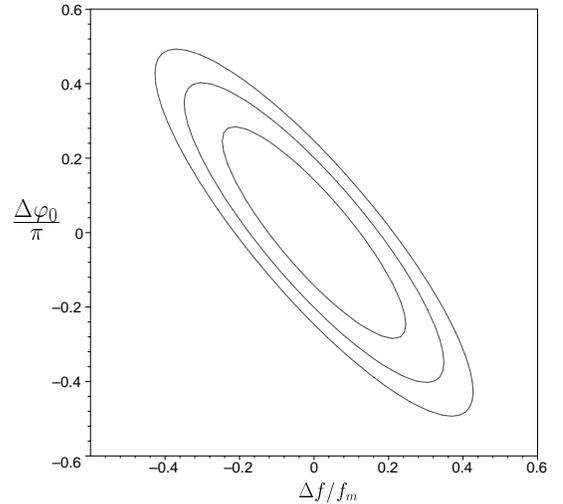}
\vspace{5mm}
\caption{The template overlap contours on the $(f,\varphi_0)$ cylinder.}
\label{overlapfphi}
\end{figure}

Setting $ d \theta = d \phi = 0$ leaves the two dimensional metric
on the $(f,\varphi_0)$ cylinder:
\begin{equation}
   ds^2_2 = \frac{2}{3}\left( \frac{\pi}{f_m} df + \frac{3}{4}
   d\varphi_0\right)^2 +\frac{1}{8} d\varphi_0^2 \, .
\end{equation}
Using this metric to plot the level sets of the overlap function, the
contours for 90\%, 80\% and 70\% overlap are shown in Figure
\ref{overlapfphi}.
One rather surprising result is that the template overlap drops very
quickly with $\Delta f$.  According to Nyquist's theorem, the
frequency resolution observations of time $T$ should equal $f_N=1/T$.
But we see from Figure \ref{overlapfphi} that the overlap drops
to 90\% for $\Delta f \sim f_N/10$. (Since we are using $T = 1$ year, it
so happens that the Nyquist frequency $f_N$ equals the modulation
frequency $f_m$.)

\begin{figure}[ht]
\vspace{80mm}
\includegraphics{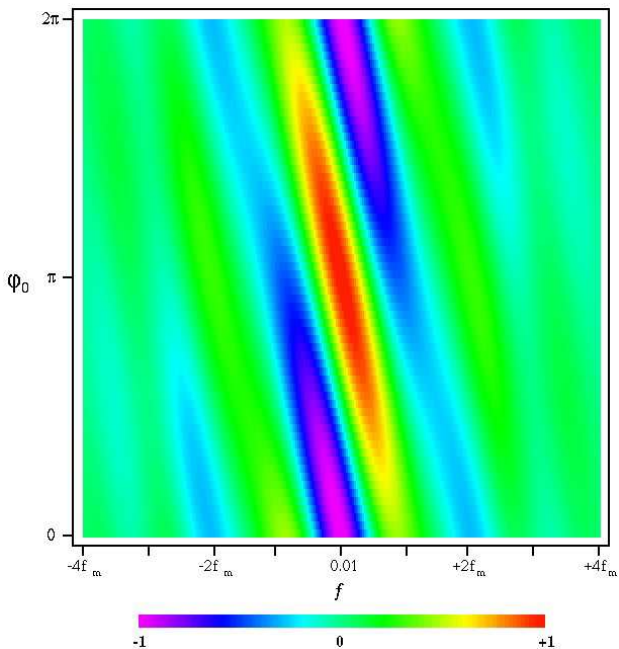}
\vspace{8mm}
\caption{The overlap of templates with all parameters equal save
frequency and orbital phase.  The reference template has
$(f,\varphi_0)=(0.01,\pi)$.  The frequency resolution was found to be
independent of the reference frequency.}
\label{numoverlapfphi}
\end{figure}

Since the metric (\ref{met}) was derived by neglecting amplitude and
phase modulation, it only gives an approximate determination of the
template overlap.  Moreover, the approximate metric neglects the
$(\imath,\psi)$ dependence completely.  In order to have a more
reliable determination of the template overlap we generated a large
template bank and studied the template overlap directly.
We see that the template overlap shown in
Figure \ref{numoverlapfphi} for $f$ vs.  $\varphi_{0}$ agrees with what
was found in Figure \ref{overlapfphi}.

\begin{figure}[ht]
\vspace{60mm}
\includegraphics{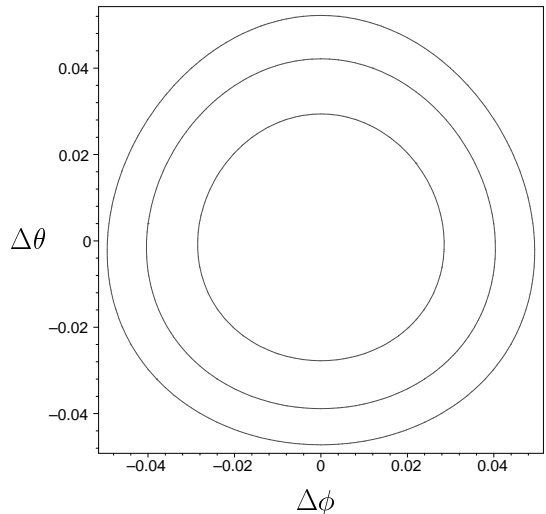}
\vspace{5mm}
\caption{The template overlap contours on the sky 2-sphere in the
neighborhood of $\theta = \pi/4 $.}
\label{overlapSky01}
\end{figure}

\begin{figure}[ht]
\vspace{43mm}
\includegraphics{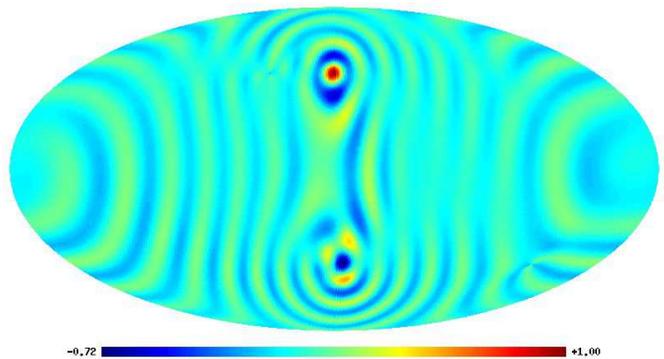}
\vspace{8mm}
\caption{The overlap of templates with all parameters equal save sky
position.  The reference template has a frequency of $f=10$ mHz and a
sky location of $(\theta,\phi)=(\pi/4,0)$.}
\label{numoverlapSky02}
\end{figure}

\begin{figure}[ht]
\vspace{60mm}
\includegraphics{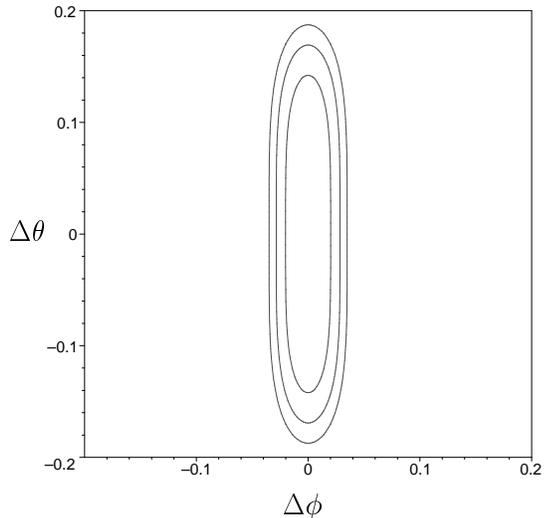}
\vspace{5mm}
\caption{The template overlap contours on the sky 2-sphere in the
neighborhood of $\theta = \pi/2 $.}
\label{overlapSky02}
\end{figure}

Setting $df = d\varphi_0 = 0$ gives us the metric on the sky 2-sphere:
\begin{equation}
   ds^2_2 = \pi^2 f^2 \left(\frac{R}{c}\right)^2 \left( \cos^2
   \theta d \theta^2 + \sin^2 \theta d \phi^2\right).
\end{equation}
It is clear from this form of the metric that the angular resolution
improves at higher frequencies, and that the metric is not that of a
round 2-sphere.  The $\cos^2\theta$ factor in front of $d \theta$
tells us that the $\theta$ resolution drops as we near the equator.
This is might seem counter-intuitive since the Doppler modulation is
maximal at the equator (it depends on $\sin \theta$).  But, the
$\theta$ resolution depends on the {\em rate of change} of the
Doppler modulation with $\theta$, which goes as $\cos \theta$.
Setting $f=10$ mHz, we plot the template overlap contours in the
neighborhood of $\theta = \pi/4 $ and $\theta = \pi/2 $ in
Figures \ref{overlapSky01} and \ref{overlapSky02} respectively.

\begin{figure}[ht]
\vspace{43mm}
\includegraphics{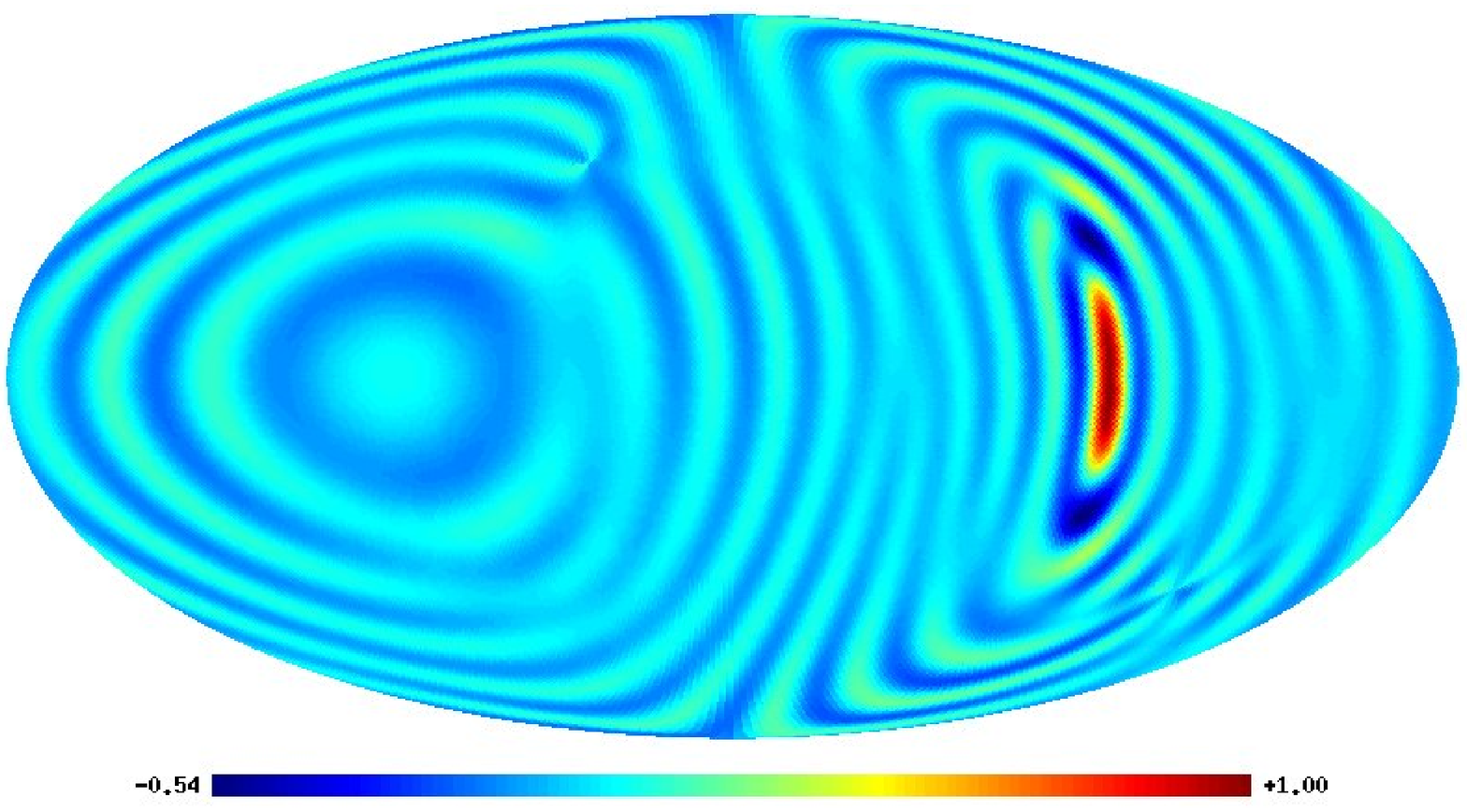}
\vspace{8mm}
\caption{The overlap of templates with all parameters equal save sky
position.  The reference template has a frequency of $f=10$ mHz and
is close to the galactic center,
$(\theta,\phi)=(1.66742,4.65723)$.}
\label{numoverlapSky01}
\end{figure}

\begin{figure}[ht]
\vspace{43mm}
\includegraphics{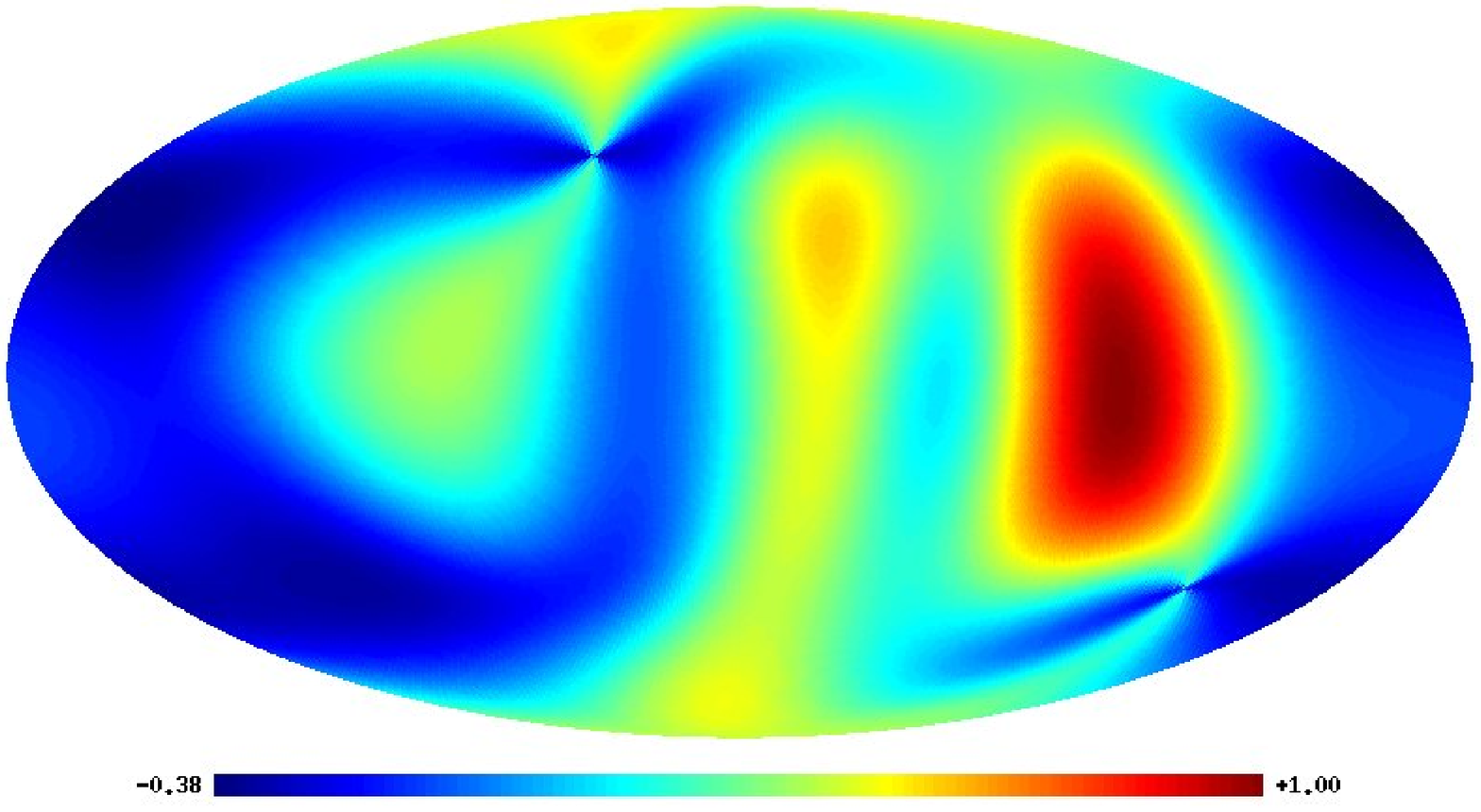}
\vspace{8mm}
\caption{The overlap of templates with all parameters equal save sky
position.  The reference template has a frequency of $f=1$ mHz and
is close to the galactic center,
$(\theta,\phi)=(1.66742,4.65723)$.}
\label{numoverlapSky03}
\end{figure}

\begin{figure}[ht]
\vspace{43mm}
\includegraphics{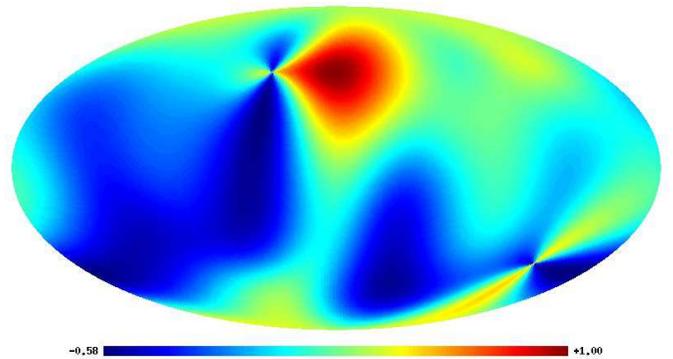}
\vspace{8mm}
\caption{The overlap of templates with all parameters equal save sky
position.  The reference template has a frequency of $f=1$ mHz and
a sky location of $(\theta,\phi)=(\pi/4,0)$.}
\label{numoverlapSky04}
\end{figure}

The template overlaps shown in Figures \ref{numoverlapSky02} and
\ref{numoverlapSky01} ($\theta$ vs.  $\phi$) agree with those in found
in Figures \ref{overlapSky01} and \ref{overlapSky02}.  The overlaps
shown in Figures \ref{numoverlapSky03} and \ref{numoverlapSky04} (also
$\theta$ vs.  $\phi$) confirm our expectation that the angular
resolution decreases with decreasing frequency.

The template overlap as a function of inclination and polarization angle
turns out to be a very sensitive function of location in parameter
space. While independent of frequency and orbital phase, the metric
functions $g_{\imath\imath}$ and $g_{\psi\psi}$ range between 0 and 326 as
$(\theta,\phi,\imath,\psi)$ are varied. Taking a uniform sample of
$1.6 \times 10^{9}$ points in $(\theta,\phi,\imath,\psi)$, we found that
$g_{\imath\imath}$ had a mean value of $0.4664$, a median value of
$0.043$, and that 90\% of all points had $g_{\imath\imath}< 1.273$.
Similarly, $g_{\psi\psi}$ had a mean value of $2.101$, a median value of
$2.0$, and that 90\% of all points had $g_{\psi\psi} < 2.251$. The
analytic expressions for $g_{\psi\psi}$, $g_{\imath\imath}$ and
$g_{\psi\imath}$ were found to agree with direct numerical calculations of
the template overlap.

\begin{figure}[ht]
\vspace{80mm}
\includegraphics{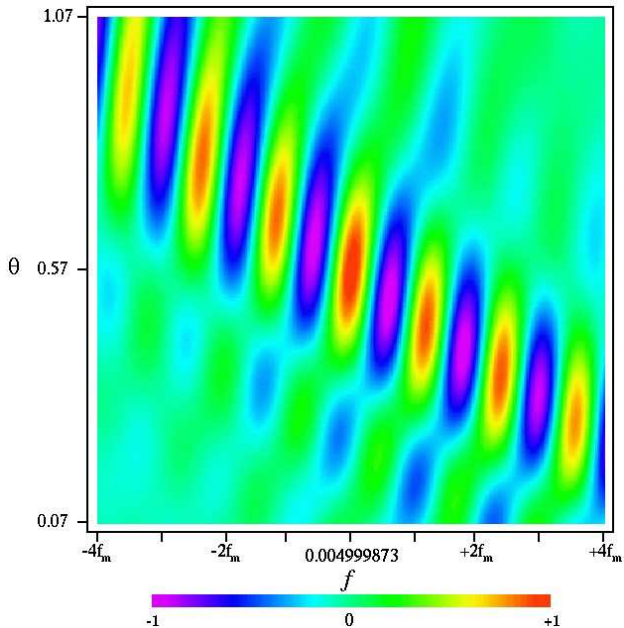}
\vspace{8mm}
\caption{An example of non-local parameter degeneracies in the $f$, $\theta$
plane.}
\label{nonlocal}
\end{figure}

\subsection{Degeneracies}\label{subsec:Degenerate}
What the metric can not tell us about are the non-local degeneracies that
occur in parameter space. A mild example of a non-local degeneracy can be
seen in Figure \ref{numoverlapSky02}, where there are secondary maxima in
the template overlap in the southern hemisphere. Physically this occurs because
the dominant Doppler modulation is unable to distinguish between sources above
and below the equator. This strong degeneracy is ameliorated by the amplitude
and phase modulations, which are sensitive to which hemisphere the source is
located in. In the course of applying the gCLEAN
procedure we discovered several other much stronger non-local parameter
degeneracies. By far the worst were those that involved frequency and
sky location. The secondary maxima sometimes had overlaps as high as 90\%.
An example of a non-local parameter degeneracy in the $f$ - $\theta$ plane
is shown in Figure \ref{nonlocal}. The reference template has $f=4.999873$ mHz,
$\theta = 0.5690$, $\phi=0.643$, $\imath = 1.57$, $\psi=0.314$, and
$\varphi_0 =0.50$. The strongest of the secondary maxima is located at
$f=4.999910$ mHz, $\theta=0.4615$, and has an overlap of 90\% with the reference
template. In principle, a sufficiently fine template grid should always find
the global maxima, but in practice, detector noise and interference from
other sources can cause gCLEAN to use templates from secondary maxima.
We will return to this issue when discussing the source identification
and reconstruction procedure.

\subsection{Counting templates}\label{subsec:CountingTemplates}
Deciding what level of template spacing is acceptable depends on two
factors: the signal-to-noise level and computing resources.  Given a
signal-to-noise level of SNR, there is no point having the template
overlap exceeding $\sim (1-1/{\rm SNR})\times 100$\%.  For the searches
described in the next section we made a trade-off between coverage and
speed, and chose template spacings that gave a minimum template
overlap of $\sim 75\%$ in each parameter direction ({\it i.e.} with
all parameters equal save the one that is varied). We chose to study
sources with frequencies near 5 mHz, and used a uniform template grid
with spacings  $\Delta f = f_m/5$, $\Delta \varphi_0 = \pi/4$, $\Delta
\theta =\Delta  \phi = 3.7^o$, $\Delta \imath = \pi/7$ and
$\Delta \psi = \pi/9$. A better approach would be to vary the template
spacing according to where the templates lie in parameter space.
As explained earlier, each set of
$5\times 4 \times 3072 \times 7 \times 9 = 3,870,720$ templates can be used
to cover a frequency range of $10^4 f_m$. At worst, a source may lie half
way between two templates, so a $\sim 75\%$ template overlap translates into
a $\sim 92\%$ source overlap. After the coarse template bank has been used to
find a best match with the data, we refine the search in the neighborhood
of the best match using templates that are spaced twice as finely in each
parameter direction.

To implement the gCLEAN procedure, a template bank was constructed by
gridding the sky using the HEALPIX hierarchical, equal area
pixelization scheme \cite{HEALpix}.  The HEALPIX centers provide sky
locations $(\theta, \phi)$ to build up families of templates
distributed across the parameters $(f, \imath, \psi, \varphi_0)$.

\section{Instrument Noise}\label{noise}
In order to construct a demonstration of the gCLEAN method, it is
necessary not only to characterize the binary signals themselves, but
also the noise in the detector.  Instrumental noise can have important
consequences for the gCLEAN process, particularly in low signal to
noise ratio situations, where random features in the noise spectrum
of the instrument could conspire to approximate the modulated signal
from a binary.

The total output of the interferometer is given by the sum of the
signal and the noise:
\begin{equation}
   h(t) = s(t) + n(t) \, .
\end{equation}
Assuming the noise is Gaussian, it can be fully characterized by
the expectation values
\begin{equation}
   \langle \tilde{n}(f) \rangle =0 \, , \quad {\rm and} \quad \langle
   \tilde{n}^*(f) \tilde{n}(f')\rangle = \frac{1}{2}\delta(f-f')S_n(f)
   \, ,
\end{equation}
where $S_n(f)$ is the one-sided noise power spectral density. It is
defined by
\begin{equation}\label{npow}
   \langle n^2(t) \rangle = \int_0^\infty df S_n(f)\ ,
\end{equation}
where the angle brackets denote an ensemble average.  The one-sided
power spectral density is related to the strain spectral density by
$S_n(f)=|\tilde{h}_n(f)|^{2}$.

Expressing the noise as a discrete Fourier transform:
\begin{equation}
   n(t) = \sum_j n_j e^{2\pi i f_m j t},
\end{equation}
a realization of the noise can be made by drawing the real and
imaginary parts of $n_j$ from a Gaussian distribution with zero mean
and standard deviation
\begin{equation}
   \sigma_j = \frac{\tilde{h}_n(f_m j)}{\sqrt{2}} \, .
\end{equation}

The signal-to-noise ratio in a gravitational wave detector is
traditionally defined as
\begin{equation}
   {\rm SNR}(f) = \sqrt{\frac{S_s(f)}{S_n(f)}} \, ,
\end{equation}
where $S_s(f)$ is the one-sided power spectral density of the
instrumental signal\footnote{$S_{s}(f)$ is the power spectral density
due to gravitational waves folded together with the instrumental
response function.}.  Given a particular set of sources, each with
their own modulation pattern, and a particular realization of the
noise, the quantity ${\rm SNR}(f)$ will vary wildly from bin to bin.
A more useful quantity is obtained by comparing the signal to noise
over some frequency interval of width $\Delta f$ centered at $f$:
\begin{equation}
   {\rm SNR}(f,\Delta f) = \sqrt{\frac{\{S_s(f)\}}{\{S_n(f)\}}} \, ,
\end{equation}
where
\begin{equation}
   \{ S(f) \} = \int_{f-\Delta f/2}^{f+\Delta f/2} S(x) \, dx\, .
\end{equation}
A good choice is to set $\Delta f$ equal to the typical bandwidth of a
source.

\section{Source identification and subtraction}\label{gclean}

The procedure for subtraction is intimately tied to the task of source
identification, as sources with overlapping bandwidths interfere with
each other. Overlapping sources have to be identified and removed in
a simultaneous, iterative procedure called the gCLEAN algorithm.

The task of gCLEAN can be understood by thinking of the LISA data
stream as an $N$ dimensional vector $\vec S$ which represents the sum
of all the sources the algorithm is seeking to subtract.  $\vec S$ is
called the {\it total source vector}.  The ideal output of the gCLEAN
algorithm is a set of basis vectors and their amplitudes ({\it i.e.},
sources $\vec s_i$) which contribute to $\vec S$:
\begin{equation}
    \vec S = \vec s_1 + \vec s_2 + \dots \ .
    \label{SSum}
\end{equation}

The basis vectors which contribute to individual sources $\vec s_i$ are
the unit-normed templates on the parameter space, ${\hat t}_{j}$,
built from Eq.\ (\ref{quick}).  In principle, the vector space of
templates will be quite large, where the number of basis vectors $M$
is much greater than the dimensionality $N$ of the source vector.

A typical application may attempt to CLEAN a frequency window of width $\Delta f$.
The source vector $\vec S$ has dimension $N= 2 \Delta f/ f_N$, where the factor of
two accounts for the real and imaginary parts of the Fourier signal.
We typically considered frequency windows of size $\Delta f
\approx 1 \; \mu {\rm Hz}$ and observation times of $T=1$ year,
so $N\approx 60$. By contrast, the number of templates used in a search
over that data stream is of order $10^{8}$.  This discrepancy in
size naturally leads to the possibility of multiple solutions,
implying that the problem is ill posed. What gCLEAN does is return a
best-fit solution in much the same vein as a singular value
decomposition.

\subsection{CLEANing}
The first step in the gCLEAN procedure is to consider the inner
product of each template ${\hat t}_{i}$ with the source vector $\vec
S$, which represents the data stream from the interferometer.  The
``best fit'' template, ${\hat t}_{j}$, is identified as the template
with the largest overlap with $\vec S$, and a small amount $\epsilon$
is subtracted off:
\begin{equation}
   \vec S' = \vec S - \epsilon (\vec S \cdot \hat t_j) \hat t_j \, ,
\end{equation}
where $(\vec S \cdot \hat t_j) \hat t_j$ is gCLEAN's best estimate of
$\vec S$.  The template ${\hat t}_{j}$ and amount removed are recorded
for later reconstruction.

The procedure is iterated until only a small fraction of the original power
remains (for the simulations presented below, the fraction was chosen to be $1\%$).
It should be emphasized that the data stream that remains after this process
is not the CLEANed data stream. By design gCLEAN will remove a pre-set fraction
of the original power, no matter what the original signal is composed of.
It is only after reconstructing the sources from the gCLEAN record that we
can meaningfully attempt to remove a source from the data stream.

The pieces which are subtracted off in the gCLEAN procedure are
assumed to be portions of individual sources $\vec s_i$, the ensemble of
which form the total signal $\vec S$; during {\it reconstruction}
these pieces are resummed into representations of the individual sources.

\subsection{Reconstruction}

The gCLEAN procedure cannot produce a perfect match with a raw data
stream from LISA, due to the discrete griding of parameter space,
the interference between the frequency components of different sources
and instrument noise. These effects serve to generate subtracted elements
which are close, but not identical to each other. The task during
reconstruction is to identify which combination of subtracted elements are
close enough together that they are considered to be manifestations of a single
source.

Reconstruction is implemented by finding the brightest element in the
list of saved matches produced by gCLEAN, and computing the overlap of
this element with all other saved elements.  For a given overlap
threshold, all sources with strong overlap are considered to be
``close'', and are summed together to represent a single source.  The
procedure is iterated over the remaining saved elements until every
element in the gCLEAN record has contributed to a source.  The frequency
and source location parameters for the reconstructed sources are taken
to be a weighted average of all the matches contributing to that
source, where the weighting is given by the individual match
amplitudes.

The procedure is complicated by the non-local parameter degeneracies
discussed in section \ref{subsec:OverlapParams}. The reconstruction may
combine contributions that are close in terms of template overlap, but
far apart in terms of the template metric. It makes no sense to average
the parameters of metrically distant templates. For this reason we only
use contributions that are metrically close when calculating the
weighted averages of the source parameters. This can lead to several
different best fit values for the reconstructed source.

We encounter an additional difficulty when trying to reconstruct the source
amplitudes $A$ and ${\cal A}$. Consider a source with amplitude $A$.
If gCLEAN performs $n$ subtractions from this source the remaining
amplitude will be
\begin{equation}
A_n \approx A (1-\epsilon)^n \, .
\end{equation}
The equation is only approximate as other reconstructed sources may
have added or subtracted from the source in question during the course
of the gCLEAN procedure. If we simply add together the $n$ contributions
identified by gCLEAN, the amplitude of the reconstructed source will
equal
\begin{equation}
A_r \approx A \left(1 - (1-\epsilon)^n \right) \, .
\end{equation}
In other words, gCLEAN will tend to under estimate the amplitude of
a source. To compensate, we multiply the initial reconstruction by
a factor of $1/\left(1 - (1-\epsilon)^n \right)$ to arrive at the
final reconstruction which gives a better estimate of $A$. Using
this estimate, along with the weighted averages for
$(\theta,\phi,\imath,\psi)$, we can use Equation \ref{amps} to
calculate ${\cal A}$. Unfortunately, any errors in the determination of
$(A,\theta,\phi,\imath,\psi)$ adversely affect our determination of
${\cal A}$. Because of this, the intrinsic amplitude of a source, ${\cal A}$,
is usually the worst determined quantity.

The reconstruction procedure usually produces more reconstructed sources than
there were sources in the input data stream. Most of these additional ``sources''
have very small amplitudes, and their existence can be attributed to detector
noise or the formation of a blended version of two or more real sources. For
this reason, we only consider reconstructed sources with an amplitude that
exceeds the noise in the detector. Occasionally gCLEAN gets confused and produces
two fits to a single source that are nearby in parameter space, but not close
enough to have been identified as one source. We discuss some ideas for getting around
this problem in Section \ref{future}.

\subsection{Isolated Sources}
\begin{figure}[ht]
\vspace{55mm}
\includegraphics{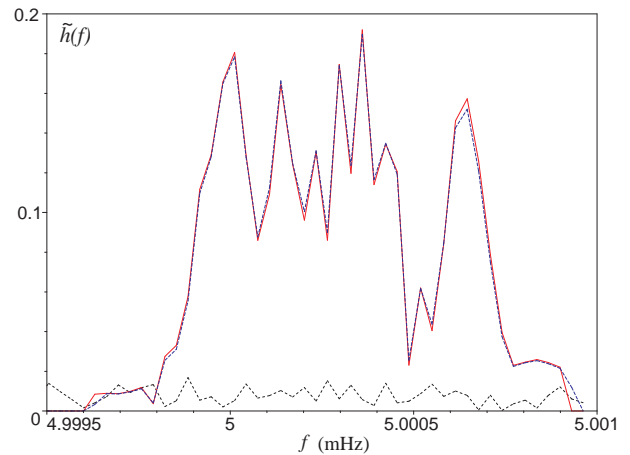}
\vspace{5mm}
\caption{The solid line is the strain spectral density of the source, the dotted line is
that of the noise and the dashed line indicates the strain spectral density of the
composite source
created by gCLEAN.}
\label{spectrum1src}
\end{figure}

Figure \ref{spectrum1src} shows the result of a gCLEAN run
carried out for an isolated source. The source parameters are listed
in table \ref{1SourceTable}. The strain spectral density of the
source and detector noise are shown in Figure \ref{spectrum1src}, along with the
composite source  built by gCLEAN. The signal to noise ratio was $9.75$ across the
bandwidth of the signal.

\begin{figure}[ht]
\vspace{55mm}
\includegraphics{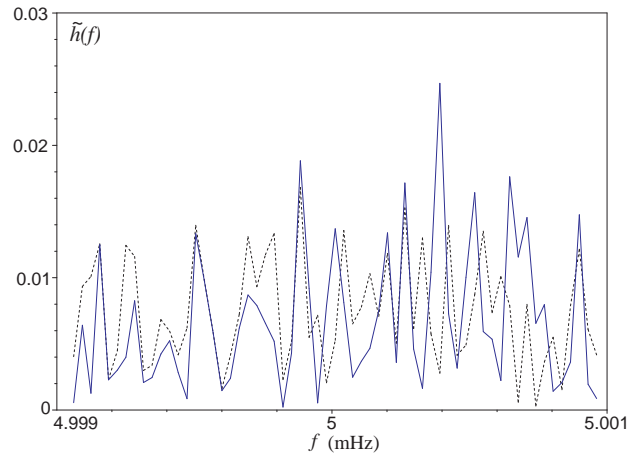}
\vspace{5mm}
\caption{The solid line is the CLEANed strain spectral density and
the dotted line is the original detector noise.}
\label{oneclean}
\end{figure}

The output from gCLEAN was then fed through the reconstruction procedure
using an overlap threshold of $0.7$. The source parameters were estimated
by taking a weighted average of the template parameters used to create the composite
source. These estimates are listed in table \ref{1SourceTable}.
The reconstruction procedure was able to fit all of the source parameters
very well save for the intrinsic amplitude ${\cal A}$. The error in
${\cal A}$ is primarily due to the error in the inclination. The large error in
${\cal A}$ translates into a large error in the distance to the source $d$.
The reconstructed parameter values for the source can be fed into our
detector response model, and the resulting strain can be subtracted from
the data stream. The CLEANed data stream is shown in Figure \ref{oneclean}.
The residual is comparable to the noise in the detector.

\begin{table}
    \centering
    \caption{The parameters for the isolated source example. The first
row lists the input values while the second row list the reconstructed values.}
    \vspace*{0.1in}
    \begin{tabular}{|c|c|c|c|c|c|c|c|}
     \colrule
    $f$ (mHz) & ${\cal A}$ & $A$ & $\theta$ & $\phi$ & $\imath $ & $\psi$ &
    $\varphi_{0}$ \\
      \colrule
    5.000281 & 0.556  & 0.648 & 0.79 & 2.21 & 2.45 & 1.62 & 0.71\\
     \colrule
     \colrule
    5.000280 & 0.786  & 0.646 & 0.79 & 2.21 & 2.11 & 1.63 & 0.81 \\
     \colrule
    \end{tabular}
    \label{1SourceTable}
\end{table}

What we have shown is that the gCLEAN procedure is able to successfully
remove isolated sources from the LISA data stream. {\em The procedure works
equally well if there are one or one million isolated sources.} The key
is that the sources are isolated, {\it i.e.} the signals do not overlap
in Fourier space. When the sources are overlapping they interfere with each
other and the CLEANing is more difficult.

\subsection{Overlapping Sources}

\begin{table}
    \centering
    \caption{The source parameters used to generate the overlapping signals.}
    \vspace*{0.1in}
    \begin{tabular}{|c|c|c|c|c|c|c|c|c|c|}
     \colrule
    $\#$ & $f$ (mHz) & SNR & ${\cal A}$ & $A$ & $\theta$ & $\phi$ & $\imath $ & $\psi$ &
    $\varphi_{0}$ \\
      \colrule
    1 & 4.999729 & 14.3 & 0.514  & 0.741 & 0.66 & 3.32 & 2.86 & 1.42 & 1.84 \\
    2 & 4.999904 & 7.7 & 0.322  & 0.399 & 2.87 & 0.26 & 2.64 & 0.26 & 2.00 \\ 
    3 & 5.000216 & 8.8 & 0.829  & 0.457 & 1.40 & 4.35 & 1.57 & 1.10 & 1.18 \\
     \colrule
    \end{tabular}
    \label{3SourceTable}
\end{table}

\begin{figure}[ht]
\vspace{55mm}
\includegraphics{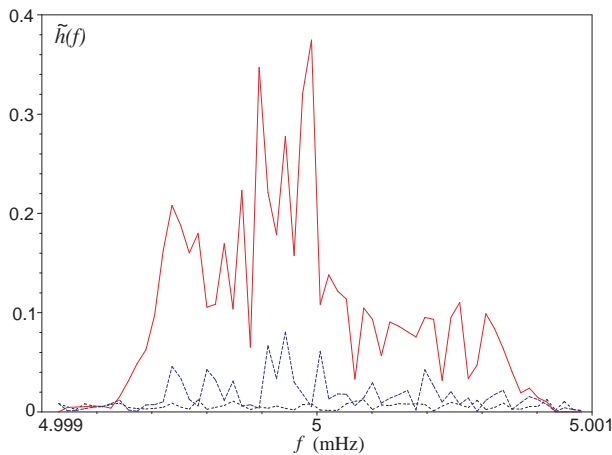}
\vspace{5mm}
\caption{The strain spectral density for the overlapping source example. The
solid line is the signal, the dashed line is the residual strain ({\it i.e.} the
CLEANed signal), and the dotted line is the detector noise.}
\label{3strain}
\end{figure}

To get a feel for how the gCLEAN procedure copes with overlapping sources, we
considered three sources with barycenter frequencies near 5 mHz that are within $\sim 5$
frequency bins of each other. The total signal to noise in the simulation was equal
to ${\rm SNR}=19.5$.
Table \ref{3SourceTable} list the randomly generated
source parameters and the signal-to-noise ratio for each source.
The modulations described in Section \ref{sec:Modulation}
cause the measured strains to overlap in frequency space. The composite strain
spectral density produced by the three sources is shown in Figure \ref{3strain},
along with the detector noise used in the simulation. Also shown is the residual
strain after the three reconstructed sources have been subtracted from the
original data stream.

The usual procedure was followed: The simulated LISA data stream
was fed into the gCLEAN algorithm, and
the output from the gCLEAN run was used to reconstruct the sources. An overlap
threshold of $0.7$ was used in the reconstruction. The reconstruction produced
five reconstructed sources with signal to noise ratios greater than one. The
reconstructed source parameters are listed in Table \ref{3ReconTable}. The first
three reconstructed sources are fair reproductions of the input sources. The
frequencies and sky locations are well determined, but there are larger errors
in the determination of the inclination, polarization angle, and orbital phase.
The strain amplitude in the detector, $A$, is fairly well determined for sources
1 and 2, but poorly determined for source 3. Once again, the intrinsic amplitude
of each source, ${\cal A}$, is the least well determined parameter.

\begin{table}
    \centering
    \caption{The reconstructed sources}
    \vspace*{0.1in}
    \begin{tabular}{|c|c|c|c|c|c|c|c|c|c|}
     \colrule
    $\#$ & $f$ (mHz) & $A_r$ & ${\cal A}$ & $A$ & $\theta$ & $\phi$ & $\imath $ & $\psi$ &
    $\varphi_{0}$ \\
      \colrule
    1 & 4.999729 & 0.646 & 0.942 & 0.731 & 0.67 & 3.33 & 1.98 & 1.10 & 1.22 \\
    2 & 4.999910 & 0.204 & 0.477 & 0.343 & 2.85 & 0.42 & 1.79 & 0.44 & 1.10 \\
    3 & 5.000214 & 0.163 & 0.543 & 0.314 & 1.50 & 4.37 & 1.57 & 1.04 & 1.49 \\
     \colrule
    4 & 5.000089 & 0.061 & 0.868 & 0.320 & 2.63 & 5.33 & 1.42 & 2.88 & 1.67 \\
    5 & 5.000336 & 0.050 & 0.217 & 0.177 & 2.36 & 4.12 & 1.98 & 0.98 & 1.78 \\
     \colrule
    \end{tabular}
    \label{3ReconTable}
\end{table}

In addition to recovering the three input sources, the reconstruction procedure
produced two spurious sources. The degree to which these sources were used by
the gCLEAN procedure is measured by the amplitude $A_r$. It is clear from the $A_r$
values that the first three reconstructed sources played a much more significant role
in the gCLEAN procedure than the two spurious sources. In terms of the amount of signal
removed during the the gCLEAN procedure, the reconstructed sources had signal
to noise ratios of $12.1$, $3.9$, $4.3$, $1.6$, and $1.1$ respectively. This suggests
that we should only consider sources with signal to noise ratios above
${\rm SNR} \sim 2$ when performing the reconstruction. Our hope is that, when we
implement some of the improvements described in Section \ref{future}, the CLEANing
procedure will produce fewer and weaker spurious reconstructions.

\begin{figure}[ht]
\vspace{130mm}
\includegraphics{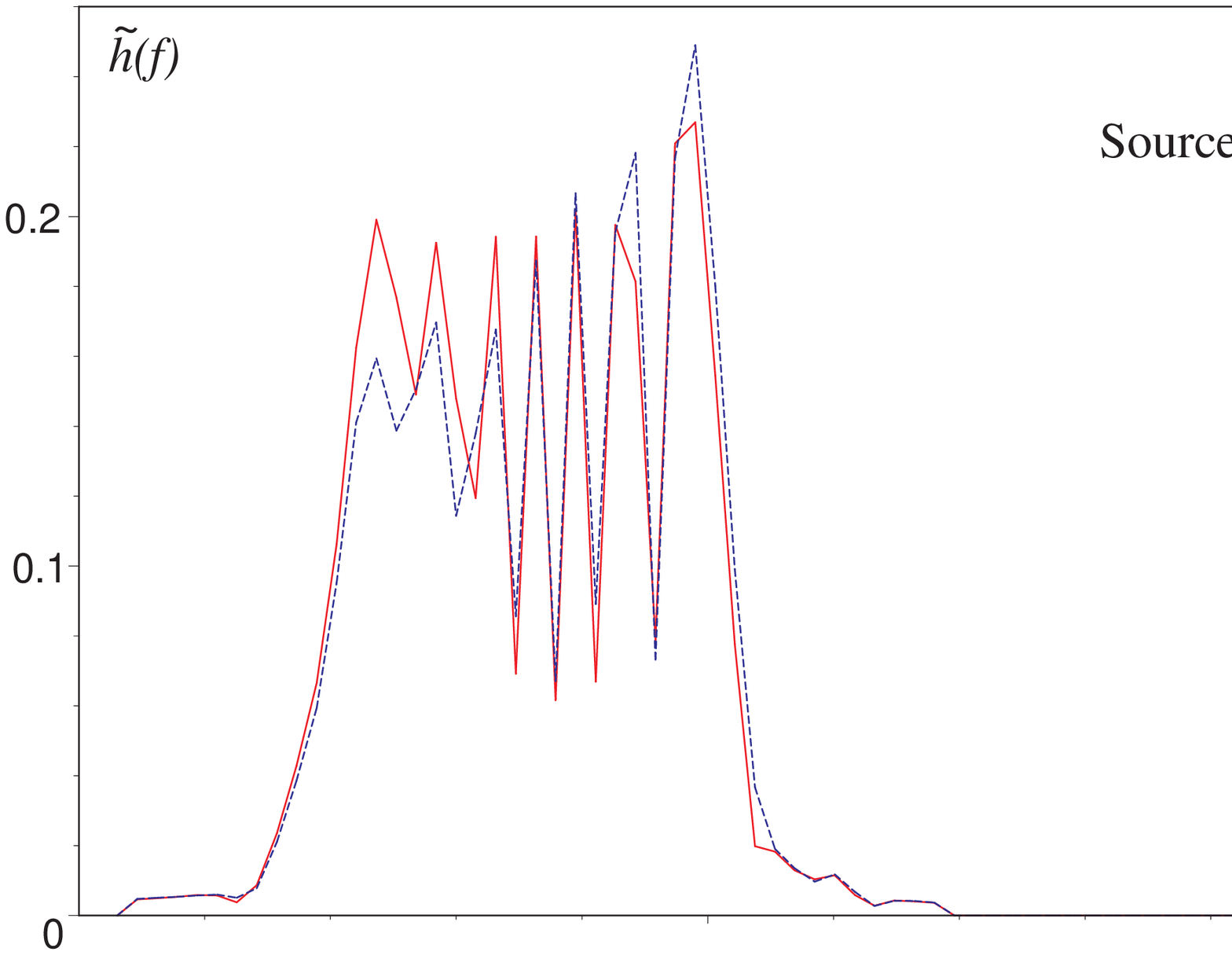}
\includegraphics{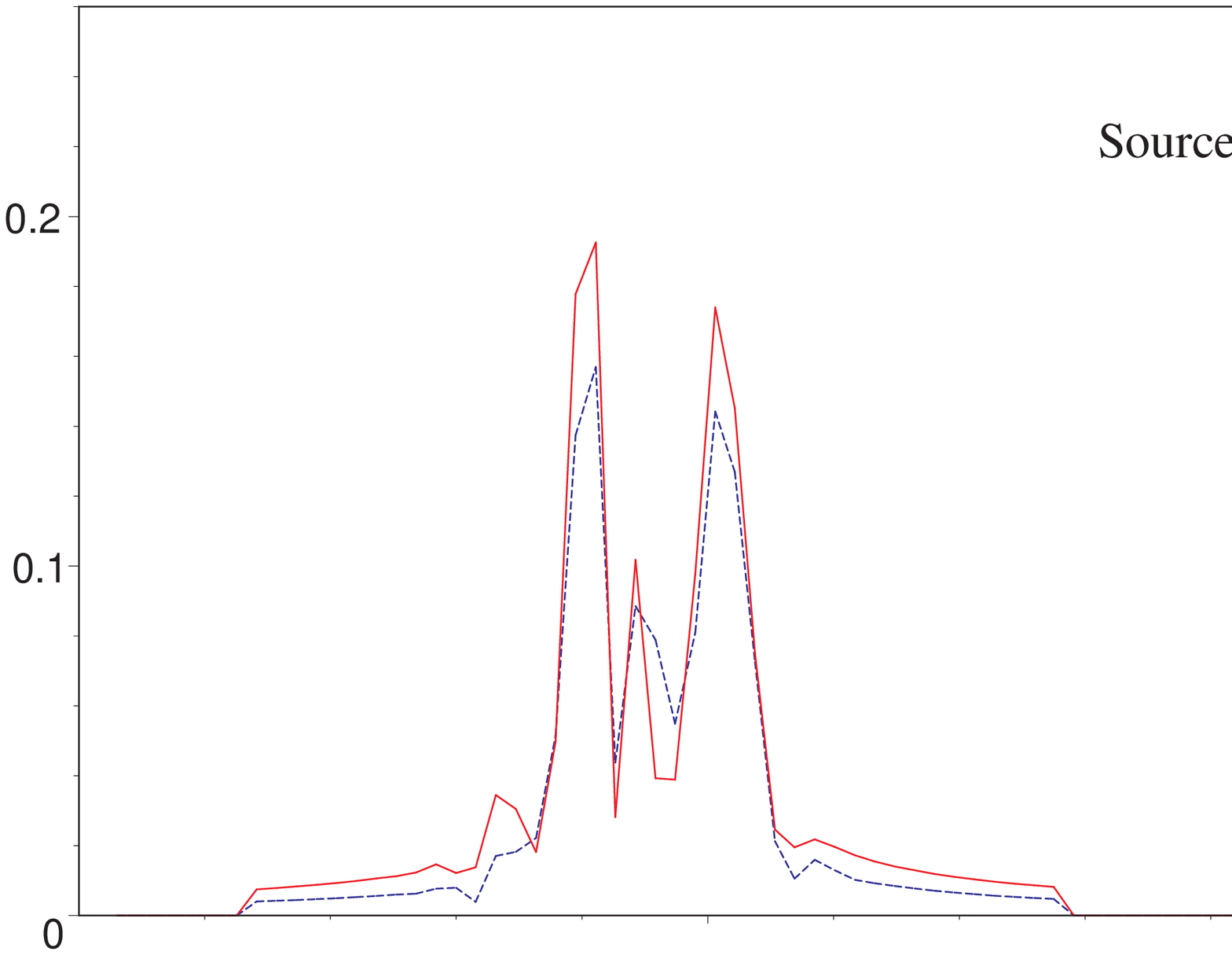}
\includegraphics{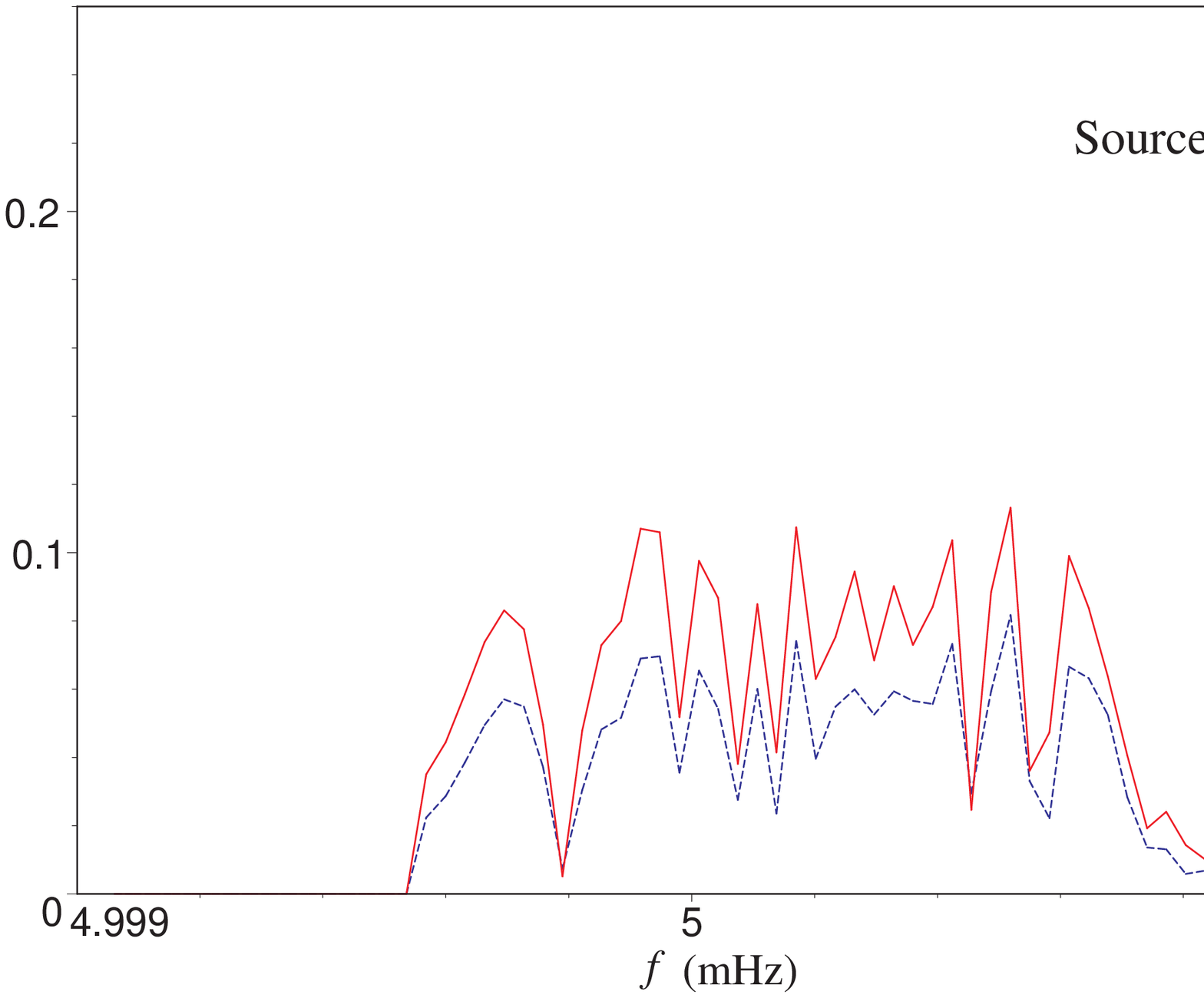}
\vspace{5mm}
\caption{The strain spectral densities of the three sources (solid lines) and
their reconstructions (dashed lines).}
\label{3recon}
\end{figure}

The strain spectral densities for the sources and their reconstructions
are shown in Figure \ref{3recon}. The reconstruction procedure
underestimates the amplitudes of sources 2 and 3. This can be attributed to
the power lost to spurious sources in the gCLEAN procedure. The CLEANed
strain spectral density shown in Figure \ref{3strain} was produced by subtracting
the three reconstructed sources shown in Figure \ref{3recon} from the input
data stream. The residual strain is down by a factor of $\sim 10$ from the
input level, but is still considerably larger than the detector noise.
The goal of future work will be to improve the
source identification and subtraction procedure to the point where multiple
sources with overlapping signals can be removed from the LISA data stream
leaving a residual that is comparable to the instrument noise.

\section{Future Work}\label{future}
The gCLEAN algorithm described here is only the first step in a
program to understand how to remove binaries from the LISA data
stream.  In particular, the limitation of the simulations presented
here are for small numbers of binaries, and at frequencies above the
expected regime where multiple overlapping binaries contribute power
in every bin of the power spectrum (this occurs at $f \simeq 3$ mHz, for
an assumed bin width of $f_N\simeq 1/{\rm year}$).

A key question is how effectively can gCLEAN 
identify binaries which have merged together to form a confusion
limited background?  While gCLEAN will subtract any signal out of the 
data stream down to a prescribed level in total power, using as many 
templates as necessary to remove the ``signal'', the real question is how
well can it identify individual sources for later removal?
Information theory predicts an ultimate bound on 
the number of binaries which can be fit out of the LISA data stream \cite{sterl}, 
and an important question is how closely can gCLEAN approach this 
optimal limit.

A great deal of work has yet to be done in the area of optimizing the
gCLEAN procedure and making it an effective tool in the LISA data
analysis arsenal; many of them are obvious extensions to the initial
foray presented in this work. Of particular interest is extending
gCLEAN to work with multiple data streams.  The design of the LISA
observatory provides three different data streams, which can be
combined in various ways \footnote{For example, in Cutler's original
treatment of the detector modulation \cite{Cutler98}, he decomposed
the LISA data channels into two overlapping interferometric signals
labeled I and II.}. As currently implemented, gCLEAN only uses a
single data stream. We are in the process of upgrading the algorithm
so that all data streams are used. It is our hope that some of the
parameter degeneracies described in Section \ref{subsec:Degenerate}
will be broken when more than one interferometer signal is used.
At the very least, we expect the parameter estimation to be improved.
It would also be interesting to see how much better the algorithm
performs if we use more than one year of observations.

The placement of templates in parameter space
is an area where improvements in efficiency can be implemented. 
Templates now are spaced for convenience ({\it e.g.}, points on the
sky are spaced on the HEALPIX centers, which are effective for
visualization), but efficient template spacing should be developed
based on the local values of the metric on the template space.

There are also several unresolved questions about the gCLEAN algorithm
and the ultimate limits of its performance on real scientific data.  
Of particular interest is how will gCLEAN perform when other signals, such
as those from supermassive black hole binaries,
are present in the data?  The research presented here has been for the
case where only circular Newtonian binaries are present in the data stream.
It is clear from the way gCLEAN is designed to work that it will indiscriminately
remove signals from a data stream; this has important implications for
how gCLEAN should be included in the approach to LISA data analysis.  
How will gCLEAN deal with chirping binaries, or signals from 
extreme-mass ratio inspirals?  Can gCLEAN be used in a sequential 
analysis strategy, where it is used to first subtract out 
monochromatic binaries before looking for other gravitational wave 
events, or do all signals have to be simultaneously gCLEANed using 
templates for each individual type of source?

There may be better ways to extract the best fit parameter values for
the reconstructed sources. Currently we use a weighted average of the
parameters that describe the templates used to build up the reconstructed
source. A better approach may be to take each reconstructed source and
use a hierarchical search through parameter space to find which set of
parameters give the best match to the reconstructed source.

Another avenue of research is devising better strategies for
accurately fitting sources.  One idea is to attempt {\it
multi-fitting}, where gCLEAN removes more than one source at a time. 
The parameter space scales as $6^{n}$, where $n$ is the number of
sources which the algorithm is attempting to subtract out at once. 
With this in mind, it is obvious that one would
have to establish initial estimates of the parameters from a standard
gCLEAN pass in order to narrow the search area of the parameter space
used in the multi-fitting.  An aspect of the subtraction enterprise which
might benefit from such a procedure is what to do with orphaned
sources which are generated during reconstruction but do not meet
the threshold requirements to be included in the final list of
identified sources.  These orphans represent subtractions on the part
of gCLEAN which arise from either fluctuations in the detector noise
which produces a close match with templates, or more commonly,
interference between the signals of multiple sources which produced a
strong match during the gCLEAN iterations.  This type of subtraction
is inevitable, as gCLEAN has no {\it a priori} way of distinguishing
interfering sources from isolated sources; it relies only on its
ability to match the current version of the data stream to its space
of templates.

There are many obvious avenues of refinement which should be pursued
in future work to develop the gCLEAN algorithm.  We are working on
several of the issues described above, and we encourage others to pursue
aspects of the problem which are of interest to them.  To aid in the
exploration of the strengths and weaknesses of the gCLEAN algorithm,
the analysis codes used to produce the results in this paper will be made
available to the scientific community through the Working Groups of
the LISA project.

\section*{Acknowledgments}

It is a pleasure to thank Tom Prince and Teviet Creighton at Caltech,
as well as the members of the Montana Gravitational Wave Astronomy
Group - Bill Hiscock, Ron Hellings, Louis Rubbo, Olivier Poujade and
Matt Benacquista - for many stimulating discussions.  The work of N.J.C.
was supported by the NASA EPSCoR program through Cooperative Agreement NCC5-579.
The work of S.L.L.was supported by LISA contract number PO 1217163.

\end{document}